\documentclass[aps,footinbib,showpacs,showkeys,preprinttightenlines,superscriptaddress,preprintnumbers] {revtex4}

\usepackage{amssymb}
\usepackage{amsfonts}
\usepackage{amsmath}
\usepackage{bm}
\usepackage{color}
\usepackage{slashed}
\usepackage{graphicx}
\usepackage{color}
\usepackage{epsfig}
\usepackage{hhline}
\usepackage{longtable}
\usepackage{multirow}
\usepackage{rotating}
\usepackage{hyperref}
\usepackage{color}

\newcommand{\dis}[1]{\begin{equation}\begin{split}#1\end{split}\end{equation}}
\newcommand{\be}{\begin{equation}}
\newcommand{\ee}{\end{equation}}
\def\bea{\begin{eqnarray}}
\def\eea{\end{eqnarray}}

\newcommand{\eq}[1]{Eq.~(\ref{#1})}

\newcommand{\bfrac}[2]{{\left(\frac{#1}{#2} \right)  }}\newcommand{\VEV}[1]{\langle #1 \rangle}

\newcommand\tev{\,{\rm TeV}}
\newcommand\gev{\,{\rm GeV}}

\newcommand{\tildeX}{{\tilde{X} }}
\newcommand{\tildeN}{{\tilde{N} }}
\newcommand{\msinglino}{m_{\tilde{X}} }
\newcommand{\mgravitino}{m_{\tilde{G}} }

\def\bea{\begin{eqnarray}}
\def\eea{\end{eqnarray}}
\definecolor{brown}{rgb}{0.5,0.2,0.0}

\begin{document}
\title{ Dark matter asymmetry in supersymmetric Dirac leptogenesis}

\author{Ki-Young Choi}
\email{kiyoung.choi@apctp.org}
 \affiliation{Asia Pacific Center for Theoretical Physics, Pohang, Gyeongbuk 790-784, Republic of Korea}
  \affiliation{Department of Physics, POSTECH, Pohang, Gyeongbuk 790-784, Republic of Korea
}

\author{Eung Jin Chun}
\email{ejchun@kias.re.kr}
\affiliation{Korea Institute for Advanced Study, Seoul 130-722, Republic of Korea}

\author{Chang Sub Shin}
\email{csshin@apctp.org}
\affiliation{Asia Pacific Center for Theoretical Physics, Pohang, Gyeongbuk 790-784, Republic of Korea}

\color{black}
\begin{abstract}
We discuss asymmetric or symmetric dark matter candidate in the supersymmetric  Dirac leptogenesis scenario. By introducing a singlet superfield coupling to right-handed neutrinos,  the overabundance problem of dark matter can be evaded and various possibilities for dark matter candidate arise. If the singlino is the lightest supersymmetric particle (LSP), it becomes naturally asymmetric dark matter. On the other hand, the right-handed sneutrino is a symmetric dark matter candidate whose relic density can be determined by the usual thermal freeze-out process. 
The conventional neutralino or gravitino LSP can be also a dark matter candidate 
as its non-thermal production from the right-handed sneutrino
can be controlled appropriately.
In our scenario, the late-decay of heavy supersymmetric particles mainly produce the right-handed sneutrino and neutrino which is harmless to the standard prediction of the big-bang nucleosynthesis.
\end{abstract}


\preprint{APCTP Pre2012-017}
\maketitle

\section{Introduction}
\label{intro}

The absence of anti matter in our Universe is one of the main questions in cosmology and
particle physics.
One of the  suggestions to generate this "matter-antimatter asymmetry" is the leptogenesis scenario which first generates lepton asymmetry through lepton number violating operators and convert it to baryon asymmetry by electroweak sphaleron transition~\cite{Fukugita:1986hr}.

The leptogenesis mechanism can be realized even without violating lepton number in the context of Dirac neutrino models~\cite{Dick:1999je}, which is called the ``Dirac leptogenesis". Some interesting variants were proposed in~\cite{Cerdeno:2006ha,Gu:2006dc,Gu:2007ps,Gu:2007mi,Chen:2011sb}.  A realistic supersymmetric model for Dirac leptogenesis was proposed in~\cite{Murayama:2002je} and studied in detail in~\cite{ Thomas:2005rs,Thomas:2006gr,Abel:2006hr}.
However the supersymmetric Dirac leptogenesis faces an imminent cosmological problem: the overabundance of right-handed (RH) sneutrinos~\cite{Chun:2008pg}.

When the lepton asymmetry in the left-handed (LH) sector is generated,
the opposite RH (s)neutrino asymmetry of the same amount arises at the same time. 
However the total number density of RH sneutrino is the same order as that of the decaying heavy fields.
Now that the RH sneutrino has only a tiny Yukawa coupling of order $10^{-13}$,  it cannot equilibrate and 
thus  the relic density is too large to be a stable dark matter (DM) candidate since its mass comes from soft supersymmetry breaking of order of the weak scale or larger. The problem persists although it can decay. Again due to the small Yukawa interaction, it decays very late to overproduce any dark matter candidate in the model. Thus, it is required to have a dilution mechanism or a very light DM candidate without changing the successful Dirac leptogenesis in the LH sector.

Some ideas have been discussed in \cite{Chun:2008pg,Thomas:2007bk}.
One can make RH sneutrinos decay before the freeze-out of a DM candidate
by introducing unconventionally large SUSY breaking terms  \cite{Chun:2008pg}.
Ref.~\cite{Thomas:2007bk} introduces a gauged $U(1)_N$ symmetry,
under which the RH neutrino superfield is charged, but the Standard Model fields are not,
and some extra fields to cancel the gauge anomalies of $U(1)_N$.
Then the RH sneutrino can be thermalized by the gauge interaction
and also it can have an appropriate DM relic density by the usual freeze-out process.

In this paper, we propose a simple mechanism to solve the overabundance problem
by introducing  a singlet field $X$  which has a lepton-number conserving Yukawa interaction
with a RH neutrino $N$, $\lambda_X X NN$, in the superpotential.
In this scenario, there arise various interesting possibilities for asymmetric or symmetric dark matter candidates.

In Section~\ref{sec:DL} the main features of Dirac leptogenesis will be briefly reviewed. In Section~\ref{sec:model} we describe our model, and in Section~\ref{sec:darkmatter} we discuss how to avoid the sneutrino overabundance problem and study various dark matter candidates in the model
such as the singlino (the fermion component of the superfield $X$), the RH sneutrino, the standard neutralino, and the gravitino. We conclude in
Section~\ref{sec:con}.

\section{Dirac leptogenesis}
\label{sec:DL}

Our main results  do not depend on a specific model of Dirac leptogenesis, however
let us adopt a concrete model of \cite{Chun:2008pg} for illustration.
To realize the Dirac leptogenesis in supersymmetric theories, we use Non-Minimal Supersymmetric Standard Model (NMSSM) which additionally contains RH neutrinos and very heavy Dirac particles $(\Phi, \Phi^c)$ whose decay produces lepton asymmetry.  The superpotential is then given by
\dis{
W &= y_u Qu^c H_u + y_d Q d^c H_d + y_e L e^c H_d + \lambda S H_u H_d + f(S) \\
&+\,   \lambda_L L  \Phi^c H_u   + \lambda_N N S\Phi  + M_\Phi \Phi \Phi^c , \label{Wchun}
}
where
the  first line corresponds to the usual NMSSM
and the second line consists of fields which are responsible for
the Dirac neutrino mass term and leptogenesis.
Here, $\lambda_L$ and $\lambda_N$ are Yukawa coupling strengths and  $M_{\Phi}$ is heavy mass much larger than the  weak scale. We suppressed the generation index in the superpotential.
Here, all the superfields added to the NMSSM sector are the SM singlets,
and thus the gauge coupling unification is not affected.
Besides the lepton number $U(1)_L$,
we introduce a Peccei-Quinn symmetry $U(1)_{\rm PQ}$ to solve the strong CP problem
and the superpotential $f(S)=\mu_S^2 S + \mu_S' S^2 + \kappa S^3$ is supposed to be induced
after the $U(1)_{\rm PQ}$ breaking \cite{bae12}.
The corresponding charges are listed in Table~\ref{charge}.
\begin{table}[t]
\begin{center}
\begin{tabular}{|c||c|c|c|c|c|c|c|c|}
\hline
 fields& $L$  & $S$  & $N$ & $\Phi$ & $\Phi^c$&  $X $  & $ X^c$\\
\hline
$U(1)_L$ & $+1$ & $0$ & $-1$& $+1$   &  $-1$& $+2$ & $ -2$\\
\hline
$U(1)_{\rm PQ}$ & $0$  & $+1$ & $-1$& $0$  &  $ 0$& $+2$ & $0$\\
 \hline
\end{tabular}
\end{center}
\caption{The assignment of  lepton number and PQ symmetry.}
\label{charge}
\end{table}
In our model, the R-parity is preserved and thus the LSP is stable.

Below the mass scale $M_\Phi$, we can integrate out $\Phi,\Phi^c$ pair to get
the effective superpotential
\bea \label{DiracMass}
\Delta W_{\rm eff}=  \frac{\lambda_L\lambda_N}{M_\Phi}LH_uNS,
\eea
which generates the Dirac neutrino mass $m_D=
 (\lambda_L\lambda_N \mu_{\rm eff}/\lambda M_\Phi)v_u\equiv y_\nu v_u$
with $v_u = \VEV{H_u^0}$ and $\mu_{\rm eff} = \lambda\langle S\rangle$.
Note that a tiny Yukawa coupling $y_\nu=m_\nu/v_u \sim 10^{-13}$ can arise from the "Dirac seesaw" with $M_\Phi$ at a high scale:
\dis{
M_\Phi \simeq 10^{10}\gev \bfrac{\VEV{S}}{\tev}\bfrac{\lambda_L\lambda_N}{10^{-6}}.
}

The decay of heavy fields $\Phi$ and $\Phi^c$ will produce the same but opposite  asymmetries,
$\epsilon_L + \epsilon_N=0$, in the final states $LH_u$ and $\bar{N}\bar{S}$, respectively.
For this,
we assume that the reheating temperature after inflation is high enough, $T_R \gtrsim M_\Phi$, so that thermal Dirac leptogenesis can occur.
Since the interactions between $L$ and $N$ is too small, the RH neutrino sector has been already decoupled from the  LH neutrino sector when they are produced from the decay. Thus the lepton number for each sector is conserved separately once they are produced, while the total lepton number asymmetry is vanishing.
Then, the non-vanishing asymmetry normalized by the entropy density in the left-handed sector, $Y_L\equiv (n_L-n_{\bar{L}})/s$,
is converted to the baryon asymmetry $Y_B$ through the sphaleron process.
Thus, it is required to have
\dis{
Y_B \approx Y_L\approx  \epsilon_L Y_\Phi \sim 10^{-10},\label{YL}
}
where $Y_\Phi$ is the out-of-equilibrium density of the heavy superfield pair $\Phi$, $\Phi^c$.

 Here appears the problem of unwanted relics~\cite{Chun:2008pg}. Independently of the details of Dirac leptogenesis models, the successful leptogenesis implies a ``large" 
abundance of the RH sneutrinos from the decay of $\Phi$ fields, $Y_N \sim Y_\Phi \gg 10^{-10}$, 
 which can be translated into the dark matter  abundance of $Y_{DM}\gg 10^{-10}$.  The present observation of the DM abundance requires 
 \dis{ \label{mYDM}
 Y_{\rm DM} \lesssim 3.7\times 10^{-12} \bfrac{100 \mbox{ GeV}}{m_{\rm DM}}
 }
 and thus (\ref{YL}) contradicts with the conventional supersymmetric dark matter mass at the weak scale or above.  In the following section, we will propose a simple solution to this problem.

\section{Model}
\label{sec:model}

Let us now introduce an additional Dirac singlet superfield  $X, X^c$  with the charge assignment in Table 1 to allow the following superpotential:
\dis{
W_X=\frac12  \lambda_X X NN  + M_X X X^c. \label{Wx}
}
This changes thermal history of the RH sneutrinos and the dark matter phenomenology.
The superfields $X(X^c)$ and $N$ are decomposed as
\bea
X^{(c)}= X^{(c)} +\sqrt{2}\theta \tilde X^{(c)} + \theta^2 F^{X^{(c)}},\quad
N= \tilde N+ \sqrt{2}\theta N + \theta^2 F^N.
\eea
Note that we have two R-parity odd particles; the``Dirac singlino", $(\tilde X, \tilde X^c)$,
and the usual RH sneutrino $\tilde N$ whose dark matter property will be discussed in the following section.

The F-term scalar potential  is
\bea
 V_F = M_X^2 |X|^2 +  \Big|\frac12\lambda_X \tilde N^2 + M_X X^c\Big|^2 + \Big|\lambda_X
 X \tilde N\Big|^2,
\eea
and the soft SUSY breaking terms are
\bea
 V_{\rm soft} = m_{X}^2 | X|^2 + m_{X^c}^2 | X^c|^2
 + m_{\tilde N}^2 |\tilde N|^2 +
\left( \frac12 A_{X}\lambda_X \,  X \tilde N^2  +  B_{X} M_X  X X^c + h.c.\right).
\label{softterm}
\eea
For positive soft mass-squareds and rather small soft parameters, $B_X$ and  $A_X$,
all scalar fields are stabilized at the origin ensuring
the lepton number conservation.
Then the RH sneutrino $\tilde N$ gets a soft mass $m_{\tilde N}$, and the mass eigenvalues of the singlet scalars $X, X^c$ are given by
\bea
m_{ X_{1,2}}^2 &=& M_X^2 + \frac{1}{2}\left(m_{ X}^2 + m_{X^c}^2 \pm
\sqrt{(m_{ X}^2-m_{  X^c}^2)^2 + 4 |B_X M_X|^2}\right).
\label{massX}
\eea
Here the mass eigenstates are defined by
\dis{
X_1 = \cos\theta \, X + \sin\theta\, e^{i\delta}\, X^{c*}, \qquad X_2 = -\sin\theta\, e^{-i\delta} \, X + \cos\theta\, X^{c*},
}
where the mixing angle $\theta$ satisfies
\dis{
\tan 2\theta = \frac{2 |B_X| M_X}{m_X^2-m_{X^c}^2}
}
with $B_X = |B_X| e^{-i\delta}$. 

The fermion Yukawa and mass terms are
\bea\label{Yukawa}
{\cal L}_{\rm Yukawa}=
 \Big(-\lambda_X \tilde N \tilde XN  - \frac12\lambda_X  XN N
- M_X \tilde X\tilde {X^c} \Big)+ h.c. .
\eea
The LH and RH neutrino sector were in thermal contact for the temperature $T_R \gtrsim M_\Phi$
through $\Phi$ field. After $\Phi$'s are decoupled, RH neutrino sector is also decoupled from the LH neutrino sector,
however $X$, $X^c$ and $N$ fields are in thermal contact with each other through Yukawa interaction in \eq{Wx}.

\section{Dark matter in Dirac leptogenesis}
\label{sec:darkmatter}

Since the R-parity is conserved in the whole superpotential given in Eqs.~(\ref{Wchun}) and (\ref{Wx}), the LSP is stable and can be a dark matter candidate.
In our model, there are several possibilities for the LSP.
We will examine each case in the following subsections.
Our discussion in the following does not depend on the specific model for the Dirac leptogenesis
used in the section~\ref{sec:DL}.

\subsection{Asymmetric singlino dark matter, $\tildeX$}

The Dirac singlino ($\tilde X, \tilde X^c$), although
 decoupled from the LH neutrino sector, is
in thermal contact with RH (s)neutrinos through the interaction term $\lambda_X X NN$ in the superpotential.  
Thus the symmetric abundance $Y_{\tilde X^{(c)}}$ and the asymmetry $Y_{\Delta X} \ll Y_{\tilde X^{(c)}}$ are generated during the thermalization process with RH sneutrino and by its decay,  $\tilde N \to \bar N \tilde X^c$ and $\tilde N^* \to N \tilde X$. 
Therefore, the Dirac singlino is a viable asymmetric dark matter candidate
for its mass around 5 GeV:
\dis{
\Omega_{\tildeX} h^2 \sim  0.1 \bfrac{\msinglino}{5\gev} \bfrac{Y_{\Delta X}}{10^{-10}},
}
if its symmetric component can be depleted sufficiently.
The interaction terms in our model \eq{Yukawa} allow the pair-annihilation of the singlino and anti-singlino, $\tilde X \tilde X^c \to N \bar N$, through the exchange of $\tilde N$ in the $t$-channel, while forbidding self-annihilation of $\tildeX \tildeX $ or $\tildeX^c \tildeX^c$.
The pair-annihilation cross section is  given by
\bea\label{AN}
\langle\sigma v\rangle_{\tilde X \tilde X^c} = \frac{|\lambda_X|^4}{32\pi} \frac{ M_X^2}
{(M_X^2+m_{\tilde N}^2)^2}.
\eea
 If this annihilation cross section is larger than the standard freeze-out cross section, $\VEV{\sigma v}_{fr} = 2.57\times 10^{-9} \gev^{-2}$, the symmetric population annihilates away and there remains only the asymmetric part.  The condition of $\langle\sigma v\rangle_{\tilde X \tilde X^c} > \VEV{\sigma v}_{fr}$ leads to the constraint
\dis{
m_{\tildeN} \lesssim 133 \gev \bfrac{\lambda_X}{1} \bfrac{M_X}{5\gev}^{1/2},
}
assuming $M_X \ll m_\tildeN$.

\medskip

The singlino can be produced also from the decay of the lightest particle in the NMSSM sector, denoted by $\tilde\chi^0_1$, through the decay chain $\tilde\chi^0_1 \to \nu  \tildeN \to 
\nu \bar N \tildeX^c$.
Since the decay process involves the small neutrino Yukawa coupling $y_\nu$,  the lifetime of $\tilde \chi^0_1$ is 
of order of
\begin{equation}\label{lifetime-chi1}
\frac{1}{\Gamma(\tilde\chi^0_1\rightarrow  \nu\tilde N)}\sim \left(\frac{y_\nu^2}{16\pi} m_{\tilde\chi^0_1}\right)^{-1}
\approx 16\sec \left(\frac{10^{-13}}{y_\nu}\right)^2\left(\frac{200\gev}{m_{\tilde\chi^0_1}}\right).
\end{equation}
Since $\tilde \chi^0_1$ decays after the singlino dark matter freeze-out as well as
the Big-Bang Nucleosynthesis (BBN), its late-time decay might produce too much non-thermal dark matter density
or spoil the standard BBN prediction~\cite{Kawasaki:2004yh,Kawasaki:2004qu,Asaka:2006fs}.
In fact, these two problems can trivially be evaded in our scenario.
First, the dark matter relic density from the non-thermal production is given by
\dis{
\Omega^{\rm NTP, \tilde\chi_1^0}_{\tildeX} h^2 =  \frac{m_{\tilde X}}{m_{\tilde\chi^0_1}}\Omega_{\tilde\chi^0_1}h^2.
\label{NTPX}
}
Considering $\Omega_{\tilde\chi^0_1}h^2 \sim {\mathcal O}(1)$ and $m_{\tilde\chi^0_1} \sim 10^{2-3}\gev$,
the non-thermal production is subdominant for $\msinglino =5\gev$.
Second,  even though the dominant decay products are neutrinos and dark matter and thus are harmless, there exist hadronic decay modes, e.g., $\tilde\chi^0_1\rightarrow \tilde N\nu q \bar q$, $\tilde\chi^0_1\rightarrow \tildeN l W^+ , \cdots$. However the branching ratio is around $\mathcal {O} (10^{-4}-10^{-3})$  and thus it is safe from the BBN constraints~\cite{Asaka:2006fs}.

The similar arguments can be applied to the NLSP gravitino whose abundance is
proportional to the reheating temperature, $T_R$. In our model, the gravitino can also follow the decay chain of $\tilde G \to X^* \tilde X \to NN \tilde X$ or $\tilde G \to N \tilde N^* \to NN \tilde X$.  Assuming the two-body decay mode is open, the gravitino decays also very late as usual:
\begin{equation}\label{lifetime-gravitino}
\frac{1}{\Gamma(\tilde G\rightarrow X^*\tilde X \mbox{ or } N\tilde N^*)}\sim
\left( m^3_{\tilde G} \over {16\pi M_P^2} \right)^{-1}
\approx 10^7 \sec \left(\frac{300\gev}{m_{\tilde G}}\right)^3.
\end{equation}
The dark matter relic density coming from the gravitino decay is given by
\cite{Bolz:2000fu}
\dis{
\Omega^{{\rm NTP},\tilde{G} }_{\tildeX} h^2 \approx 0.0015 \bfrac{T_R}{10^{10}\gev}\bfrac{m_\tildeX}{5\gev} \bfrac{300\gev}{\mgravitino}^2\bfrac{m_{\tilde g}}{1\tev}^2.
}
Thus this contribution is also subdominant as far as  $T_R\lesssim 10^{12}\gev$.

\subsection{Symmetric RH sneutrino dark matter, $\tildeN$}

When the RH sneutrino is the LSP,  it is also a good dark matter candidate in our model.  Recall that the RH sneutrino whose mass is in the range of $10^2-10^3$ GeV causes the overabundance problem with its large symmetric and asymmetric abundance. 
However the Yukawa couplings in Eq.~(\ref{Yukawa}) allow the self- and pair-annihilation of the RH sneutrino and anti-sneutrino:
$\tilde N\tilde N\rightarrow N N$, $\tilde N^*\tilde N^*\rightarrow \bar N \bar N$,
and $\tilde N \tilde N^* \to N\bar N$ through which the RH sneutrinos are thermalized and the relic abundance is determined by the usual freeze-out.
The self-annihilation cross-sections are given by
\dis{ \label{NN-NbNb}
\VEV{\sigma v}_{\tildeN {\tildeN}}=\VEV{\sigma v}_{\tildeN^* {\tildeN}^*}= \frac{|\lambda_X|^4}{16\pi}   \frac{\left| A_X\left\{4m_\tildeN^2 - \frac12 (m_{X_1}^2+m_{X_2}^2)
+  \frac12 (m_X^2 - m_{X^c}^2)  \right\}+ B_X M_X^2\right|^2}{(4m_\tildeN^2 - m_{X_1}^2)^2(4m_\tildeN^2- m_{X_2}^2)^2},
}
where $m_X$ and $m_{X^c}$ are the soft mass defined in~\eq{softterm} and $m_{X_1}, m_{X_2}$ are the masses of the scalar $X$ given in~\eq{massX}.
The pair-annihilation cross section is given by
\dis{ \label{NNb}
\VEV{\sigma v}_{\tildeN {\tildeN}^*}= \frac{|\lambda_X|^4}{8\pi}\frac{m_{\tildeN}^2}{(m_{\tildeN}^2 + M_X^2)^2}
\bfrac{T}{m_{\tilde N}}.
}
This is the p-wave contribution and the s-wave is negligible due to the helicity suppression.
To deplete the large symmetric and asymmetric population, we require
$\VEV{\sigma v}_{\tildeN\tildeN}=\VEV{\sigma v}_{\tildeN^*\tildeN^*} \gtrsim \VEV{\sigma v}_{fr}$, which leads to the symmetric abundance of the RH sneutrino.


\medskip

Similarly to the discussions in the previous subsection, the
late-decay of the usual neutralino LSP in the NMSSM sector or the gravitino contributes to the dark matter relic density, which does not cause a trouble with the BBN.
 Repeating the previous discussion, we get the non-thermal
RH sneutrino relic density from the neutralino LSP decay:
\bea
\Omega_{\tilde N}^{\rm NTP, {\tilde\chi^0_1}}h^2 = \frac{m_{\tilde N}}{m_{\tilde\chi^0_1}}\Omega_{\tilde\chi^0_1}h^2.
\eea
The non-thermal contribution to the RH sneutrino abundance from the gravitino decay is given by
\dis{
\Omega^{{\rm NTP},\tilde{G} }_{\tildeN} h^2 \approx 0.03 \bfrac{T_R}{10^{10}\gev}\bfrac{m_\tildeN}{100\gev} \bfrac{300\gev}{\mgravitino}^2\bfrac{m_{\tilde g}}{1\tev}^2,
}
which requires $T_R \lesssim 3\times10^{10}$ GeV for the given choice of the parameters.

\subsection{Neutralino or Gravitino LSP as dark matter}

The usual lightest neutralino in the NMSSM sector or the gravitino is also a dark matter candidate 
as far as the overabundance problem, caused by the non-thermal production from the late-decays of 
abundant $\tilde X$ or $\tilde N$, can be evaded. 
It can be easily achieved in our scenario if we allow a fast decay of $\tilde X \to  \tilde N^*\bar N$ and efficient self-annihilations of $\tildeN^{(*)} \tildeN^{(*)}$ in \eq{NN-NbNb} which depletes the abundance and thus can lead to
\bea
\Omega_{\rm LSP}^{{\rm NTP}, \tilde N}h^2 = \frac{m_{\rm LSP}}{m_{\tilde N}}\Omega_{\tilde N}h^2
\lesssim 1 \,.
\eea

When the gravitino $\tilde G$ is the LSP, the usual neutralino decay to the LSP
through, e.g., $\tilde \chi^0_1 \to \gamma \tilde G$, would spoil the BBN prediction.
This problem can be also evaded by allowing the decay channel
$\tilde \chi^0_1 \to  \tilde N\bar N$ which is much faster than the above dangerous channel:
\bea
\Gamma(\tilde\chi^0_1\rightarrow \tilde N \bar N)\sim
 N_{14}^2\frac{y_\nu^2}{16\pi} m_{\tilde\chi^0_1}
\gg
\Gamma(\tilde\chi^0_1\rightarrow \gamma \tilde G) \sim
\frac{m_{\tilde\chi^0_1}^5}{48\pi M_{\rm P}^2m_{\tilde G}^2} ,
\eea
where $N_{14}$ denotes the $\tilde \chi^0_1$ component in the Higgsino $\tilde H_u$.

On the other hand, if the usual neutralino $\tilde \chi^0_1$ is the LSP,
the overabundant gravitino decay to the LSP
through, e.g., $\tilde G \to \gamma \tilde \chi^0_1$ would again spoil the BBN prediction. In our case, this problem cannot be evaded by allowing the decay channel $\tilde G \to  \tilde N\bar N$ as we have 
$
\Gamma(\tilde G \rightarrow \tilde N\bar N)\sim
\Gamma(\tilde G\rightarrow \gamma \tilde \chi^0_1)$. 
To solve this problem we have to resort to the usual solution of gravitino problem by lowering reheating temperature or increasing the gravitino mass~\cite{Kawasaki:2004yh,Kawasaki:2004qu}.

\section{Conclusion}
\label{sec:con}
We studied the possible dark matter candidates in the supersymmetric Dirac leptogenesis scenario which avoids the RH sneutrino overabundance problem.
By introducing a  singlet field $X$ coupling to the RH neutrino, we show that the RH sneutrino can be thermalized or decay to a light singlino dark matter.
If the singlino is the LSP, it is a natural asymmetric dark matter candidate whose abundance is directly connected to the baryon asymmetry of the Universe. For the RH sneutrino LSP, its relic density can be determined by thermal freeze-out of the self-annihilation ($\tilde N \tilde N \to N N$ and $\tilde N^* \tilde N^* \to \bar N \bar N$) leading to symmetric population.
In our scenario the conventional neutralino or the gravitino LSP can  also be a dark matter candidate as the RH sneutrino population can be depleted by the above process reducing the non-thermal production of the LSP.  It is also shown that the potential problem of the
late-decay of heavy supersymmetric particles spoiling the BBN prediction can be easily evaded
by making them decay to the RH neutrino and sneutrino.

\section*{Acknowledgments}

K.-Y.C and C.S.S were supported by Basic Science Research Program through the National Research Foundation of Korea (NRF) funded by the Ministry of Education, Science and Technology (No. 2011-0011083). E.J.C was supported by the National Research Foundation of Korea (NRF) grant funded by the Korea government (MEST) (No.~20120001177).
K.-Y.C and C.S.S acknowledge the Max Planck Society (MPG), the Korea Ministry of
Education, Science and Technology (MEST), Gyeongsangbuk-Do and Pohang
City for the support of the Independent Junior Research Group at the Asia Pacific
Center for Theoretical Physics (APCTP).




\end{document}